# Measurement of the Refractive Index and Attenuation Length of Liquid Xenon for its Scintillation Light

V. N. Solovov,[1] V. Chepel,[1] M. I. Lopes[*],[1] A. Hitachi,[2,1]

R. Ferreira Marques,[1] and A.J.P.L. Policarpo[1]

[1]*LIP-Coimbra and CFRM of Physics Department,*

*University of Coimbra, P–3004–516 Coimbra, Portugal*

[2]*Kochi Medical School, Nankokushi, Kochi 783-8505, Japan*

## Abstract

The attenuation length and refractive index of liquid xenon for its intrinsic scintillation light ($\lambda = 178\,\mathrm{nm}$) have been measured in a single experiment. The value obtained for the attenuation length is $364 \pm 18\,\mathrm{mm}$. The refractive index is found to be $1.69 \pm 0.02$. Both values were measured at a temperature of $170 \pm 1\,\mathrm{K}$.

PACS numbers: 78.20.Ci, 29.40.Mc, 78.35.+c

Keywords: liquid xenon, refractive index, vacuum ultraviolet, scintillation

[*] Corresponding author. *E-mail address:* isabel@lipc.fis.uc.pt

## I. INTRODUCTION

The scintillation light of liquid xenon occurs in the vacuum ultraviolet (VUV), with a single band centred at $\lambda$=178 nm and full width at half maximum (FWHM) of about $\Delta\lambda = 14$ nm [1]. The refractive index of liquid xenon for its scintillation, as well as the attenuation length of this light in the medium, are quantities of prime importance from both the practical and theoretical points of view.

Liquid xenon has been proposed as a working medium for several large radiation detectors based on the scintillation induced by radiation. Some of those proposals are at the time of writing at the stage of research and development. Several prototypes have been built and are being tested. Detailed information on this issue can be found somewhere else in the literature (Refs. 2, 3 and references therein). For the development of those detectors, the knowledge of both the refractive index of liquid xenon, $n$, and the attenuation length for the scintillation light in the medium, $L$, is of crucial importance as they influence the scintillation photon transport in the liquid and thus the performance of the detector [4]. The existing data on the refractive index of liquid xenon in the VUV region are scarce. To our knowledge, the only experimental determination of $n$ for xenon light, is that reported in Ref. 5. The assessment to the real accuracy of the data presents difficulties, as the authors do not provide enough details on the experiment. In addition to that value, there are measurements of $n$ in liquid xenon at wavelengths between 361 nm and 644 nm, for a range of temperature from 178 K down to the triple-point [6, 7]. Some authors have extrapolated these values in the visible to the UV, namely to the wavelength of the xenon scintillation, but the result is very sensitive to the functional form used for the extrapolation, which leads to values with very large uncertainties. This aspect will be addressed in more detail in Sec. V. As regards the attenuation length of scintillation (also referred to as the mean free path of scintillation photons), liquid xenon itself should be transparent to its scintillation light. The excimer emission originates in the transition from the self-trapped exciton states $^1\Sigma_u^+$ and $^3\Sigma_u^+$ to the repulsive ground state $^1\Sigma_g^+$. The free exciton absorption band is $\approx$1.3 eV above the peak of the emission band whose width is $\approx 0.6$ eV [8]. On the other hand, it is well known that small amounts of impurities can be very effective in absorbing the light. Therefore, during a long time, the observed attenuation in liquid xenon was attributed to absorption by impurities. However, in spite of the great progress achieved in purifying the liquid assessed usually by the measurement of lifetime of free electrons, various authors have reported similar and relatively short values of the photon attenuation length (about 30 cm) almost independently of purification processes, apparatuses and measuring methods [2, 9]. This fact has led to the suggestion that the attenuation length of scintillation measured in liquid rare gases is mainly determined by Rayleigh scattering [9], recently supported by theoretical calculations considering scattering of the light on density fluctuations [10]. The calculated values are in reasonable agreement with measurements both for pure liquefied rare gases and their mixtures, although the associated uncertainties are quite large.

It is worth to mention that the experimental determination of the attenuation length of photons in the liquid leads very easily to overestimated values [9, 11]. Most of the methods reported in the literature are based on the measurement of the attenuation of the light after having crossed a known (and usually variable) thickness of liquid. In most experiments it is assumed that only direct light is detected and the contribution from the reflected light is negligible. Thus, the set-up has to be designed very carefully in order to fulfil that requirement, which has been realised not to be easy [9, 11].

Concerning the theoretical point of view, the precise knowledge of the refractive index is remarkably important, among the optical properties, not only by itself but also because it has been used in evaluating the success of theoretical models and accurate calculation of other properties, namely the density and polarizability of non-polar fluids.

In this paper, we report on simultaneous measurements of the attenuation length and the refractive index of liquid xenon for its own scintillation at the temperature 170 K. The paper is organized as follows. In Sec. II the experimental method is explained, while the experimental set-up and procedure are described in Sec. III. The data analysis and results are presented in Sec. IV and discussed in Sec. V.

## II. THE EXPERIMENTAL METHOD

The experimental method used for measuring simultaneously the attenuation length of scintillation light and the refraction index of liquid xenon is schematically depicted in Fig. 1. A source of scintillation light pulses is aligned with the axis of a photomultiplier tube (PMT) at a distance $H$ from the photocathode of radius $r$. The source is inside the liquid with refractive index $n$, while the PMT is in the gas (refractive index $\approx 1$). The surface of the liquid is at a distance $h$ above the light source and parallel to the PMT window. Then, the average number of photoelectrons produced in the PMT per scintillation pulse may be written as

$$N = N_0(1+k)T_1 T_2 Q \frac{\Omega}{4\pi} \exp\left(-\frac{h}{L}\right), \qquad (1)$$

where $N_0$ is the average number of photons emitted into $4\pi$ per scintillation pulse, $T_1$ and $T_2$ are the transmitance of the liquid-gas and gas-PMT window boundaries respectively, $Q$ is the quantum efficiency of the PMT photocathode, $\Omega$ is the solid angle and $L$ is the attenuation length of scintillation light (also referred to as mean free path of the photons) in the liquid. In the case of existence of both absorption and Rayleigh scattering, to which are associated mean free paths $L_a$ and $L_s$ respectively, one has

$$\frac{1}{L} = \frac{1}{L_a} + \frac{1}{L_s}. \qquad (2)$$

In Eq. (1), $k$ is the ratio of the indirect to direct photons collected at the PMT. By indirect photons one means those photons that are collected at the PMT after undergoing Rayleigh scattering or/and reflection on the surrounding materials. Direct photons are those that did not suffer either Rayleigh scattering or reflection before being collected at the PMT. Hence, $k$ may depend on $h$, i.e., $k = k(h)$.

From Fig. 1, one sees that if $H \gg r$, then $\alpha$ and $\alpha'$ are small so that $\tan \alpha \approx \alpha$ and $\tan \alpha' \approx \alpha'$. Hence, one can write

$$r \approx \alpha h + n\alpha(H - h) \qquad (3)$$

$$\alpha \approx \frac{r}{Hn - h(n-1)} \qquad (4)$$

$$\Omega = 2\pi(1 - \cos \alpha) \approx \pi \alpha^2 \approx \pi \left(\frac{r}{Hn - h(n-1)}\right)^2 \qquad (5)$$

Considering that for small incident angles one has

$$T_1 \approx 1 - \left(\frac{n-1}{n+1}\right)^2, \qquad (6)$$

from Eqs. (1), (5) and (6), one obtains

$$N(h) = A(1 + k(h))\left(\frac{r/2}{Hn - h(n-1)}\right)^2 \left(1 - \left(\frac{n-1}{n+1}\right)^2\right) \exp\left(-\frac{h}{L}\right), \qquad (7)$$

where $A = N_0 T_2 Q$, which is constant with respect to $h$. Therefore, $n$ and $L$ can be determined by measuring $N$ as a function of $h$ and fitting Eq. (7) to the data, provided $k(h)$ is known. The computation of $k(h)$ is discussed in detail in section IV.

In the present work, measurements of $N(h)$ were carried out for two different values of $H$ so that one has two sets of data taken with the same sample of liquid but in someway different geometries.

In order to enhance the sensitivity of this method to $n$, and $L$ and thus provide accurate determination of those quantities, the dependence of $N$ on $h$ has to be dominated by the exponential and by the factors dependent on $n$. For that, $k(h)$ has to be $\ll 1$, i.e. the conditions of light collection have to be such that the amount of indirect light collected at the PMT is minimal. This was achieved with a proper design of the liquid container, as described in the next section.

## III. EXPERIMENTAL SET-UP AND PROCEDURE

A schematic drawing of the experimental set-up is presented in Fig. 2. The whole system is mounted inside a vacuum-tight cylindrical chamber (1) made of stainless steel. The scintillation light is induced by $\alpha$-particles from an $^{241}$Am source, 8 mm in diameter, deposited on a stainless steel substrate (4). The scintillation light is seen by a photomultiplier (2) (Hamamatsu R1668) with UV grade fused silica window and a 25 mm diameter photocathode, placed 230 mm above the bottom of the chamber.

The source is mounted on top of a magnesium cylinder (3), which also has a ferromagnetic stainless steel disk (5) screwed to its bottom. As the density of magnesium is lower than that of liquid xenon, the whole assembly can float on the liquid surface. Hence, it is possible to have the source in two different positions: on the bottom of the chamber, holding the source assembly by a permanent magnet (6) placed outside the chamber, and in an intermediate position between the bottom of the chamber and the PMT with the source floating in the liquid and hold in place by a diaphragm (see Fig. 2). In these two positions, the distance between the source and the PMT window, $H$, is 187 mm and 93 mm, respectively.

A stack of stainless steel diaphragms (7) is intended to minimize the amount of indirect light reaching the photocathode. The diaphragms below and above the intermediate position of the source have diameters of 36 mm and 25 mm, respectively, and their edges were shaped as shown in Fig. 2.

All the parts were washed in ultrasonic bath before assembly. After closing, the chamber was heated up to 353 K and pumped out for several days until the pressure decreased to $\sim 10^{-6}$ mbar. Next, xenon gas was circulated during several days through the chamber and the Oxisorb purifier.

The xenon gas passed through the Oxisorb purifier before being condensed in the chamber.

The level of liquid xenon was measured by means of a capacitance transducer made of a stainless steel rod (8) of 3 mm diameter inserted coaxially into a stainless steel tube (9) with inner diameter of 6 mm. The precision of the level measurement was better than 1 mm resulting from an accuracy of the transducer better than 0.1 pF.

The chamber was placed inside a dewar with alcohol cooled by liquid nitrogen to approximately 170 K (10). Two platinum thermoresistors (11) are attached to the chamber, one at the bottom and the other at the level of the PMT, allowing to monitor the temperature with a precision of 1 K.

For measuring the mean number of photoelectrons produced in the PMT per $\alpha$-particle, the output signal of the PMT was amplified by a charge sensitive preamplifier (Canberra 2005) followed by a spectroscopy amplifier (Canberra 2021) before being fed into a PC based multichannel analyzer for pulse height analysis. At the amplifier, a semigaussian shaping with a time constant of 1 $\mu$s was employed. For converting the pulse height due to $\alpha$-particles into number of photoelectrons, the location of the single-photoelectron peak in the pulse height spectrum was used. The single-photoelectron peak was due to the scintillation induced by the 60 keV $\gamma$-rays also emitted by the $^{241}$Am source. The position of single-photoelectron peak was permanently monitored to check the stability of the PMT gain. The linearity and stability of the electronics was checked by injecting pulses of variable amplitude, from a precision pulse generator, into the preamplifier input stage, several times during the course of the measurements.

The data-taking procedure was as follows. Initially, the floater with the source lied on the bottom of the chamber. The magnet, placed outside the chamber, just below its bottom, kept the floater in this position while xenon was condensed into the chamber. The condensation was performed in steps, the level of liquid increasing by 10 to 15 mm each time. After temperature stabilization, which typically took about 30 minutes, the difference between the readings of the two thermoresistors was about 2-3 K, depending on the liquid level. After that, the mean amplitude of the PMT signal obtained from the pulse height spectrum, and the level of liquid xenon were recorded every minute during 10 minutes. The procedure was repeated until the liquid had reached the surface of the PMT which was recognized by an abrupt change of the signal amplitude. In order to obtain a set of measurements of $N(h)$ with the source in the upper position (shown in Fig. 2 in dashed lines), the magnet was removed slowly. A second stepwise procedure was then followed, with the difference that, in each step, the level of liquid was decreased by allowing some amount of liquid to evaporate. The two sets of measurements were repeated several times.

The experimental results are plotted in Fig. 3. At each value of $h$, the mean value of the readings of the peak position in 10 spectra accumulated sequentially was taken as the best estimate of $N$. The standard deviation of that set of readings was taken as the uncertainty in $N$. The uncertainty in the data points ranges from about 0.5% to 1%, with the exception of the data point corresponding to the smallest value of $h$ and the source at the bottom of the chamber ($H = 187$ mm) for which the uncertainty is 1.5%.

## IV. DATA ANALYSIS

In order to determine the refractive index, $n$, and the attenuation length, $L$, Eq. (7) was fitted to the experimental data on $N(h)$, the number of photoelectrons detected per $\alpha$-particle. The difficulty of the fit lies in the fact that the parameter $k$ can also depend on the adjusted parameters $n$ and $L$, as well as on other unknown quantities involved in the transport of the light in the chamber, such as the reflectivity of the chamber materials and the probability of photon scattering in liquid xenon. This dependence can not be expressed in a simple analytical form, and therefore, it can only be assessed numerically by Monte-Carlo simulation. As every intermediate step of the fit requires the knowledge of $k$ for every experimental point at the current values of the adjustable parameters, its calculation by simulation during the fitting procedure would be extremely time consuming. To overcome this problem, the Monte-Carlo simulation was decoupled from the fitting procedure in the following way. First, for each experimental value of $h$ and a finite set of the input parameters, a detailed Monte-Carlo simulation of scintillation light propagation in the chamber was carried out, producing a kind of look-up-table with information that allows to compute the value of $k$ for any set of values of the parameters at the fitting stage, without the need of running the simulation program. In the next two subsections, this procedure is described in detail.

### A. Monte-Carlo simulation

The chamber was represented by a cylindrical volume where all the rings were included, as well as the source support. Details concerning the shape of the rings were also taken into account.

The emission of $\alpha$-particles from the source was assumed to be uniform over the surface of the source and isotropic in $2\pi$. As the range of an $\alpha$-particle in liquid xenon is small compared with the dimensions of the chamber (for $E_\alpha = 5.49\,\mathrm{MeV}$ and $T = 170\,\mathrm{K}$, this range is $\sim 50\,\mu\mathrm{m}$), it was neglected and the scintillation photons were simulated as being emitted isotropically from the emission point of the $\alpha$-particle.

Concerning the transport of the scintillation photons, each photon was followed, using the ray-tracing technique, until it was either collected at the PMT or absorbed, either in the liquid or at any surface of the chamber. In the liquid, the photon transport was simulated as follows: 1) At each photon location, the distance to the next interaction point in the liquid, $d_i$, is computed according to the exponential distribution

$$d_i = -L\log(\xi), \qquad (8)$$

$\xi$ being a random number uniformly distributed in [0,1]; 2) The distance, $d_b$, from the present location to the nearest boundary along the photon direction was calculated. 3) The photon was transported along the present photon direction by the distance $d$ equal to $d_i$ or $d_b$ whichever is smaller. 4) If $d = d_i$, then an interaction in the liquid took place, else a boundary surface was reached. In the gas, the attenuation of scintillation light was neglected.

In the liquid, the interaction processes considered for the photons were the Rayleigh scattering and the absorption, with associated mean free paths $L_s$ and $L_a$, respectively. These processes occur with probabilities $L/L_s$ and $L/L_a = 1 - L/L_s$, respectively, $L$ being related with $L_a$ and $L_s$ by Eq. (2). Henceforth, the probability of occurrence of Rayleigh scattering, $L/L_s$, will be referred to as $\eta$ for the sake of simplicity.

In the case of occurrence of an interaction in the liquid, the type of the interaction was sampled according to the probabilities referred above. In the case of scattering, the new photon direction was generated according to the cross section of the process [12], with a probability distribution function given by

$$dP = \frac{3}{8}\left(1 + \cos^2\theta\right)\sin\theta d\theta \frac{d\varphi}{2\pi} \qquad (9)$$

$\theta$ and $\varphi$ being the polar and azimuthal angles relative to the photon initial direction. The photon energy is unchanged and its transport continues.

The processes taken into account at each boundary and their probabilities are listed in Table I. Regarding any metal surface (wall, ring or source support surface), reflection, specular as well as diffuse, and absorption were considered. The probabilities of these processes were taken equal to $R*m$, $R*(m-1)$ and $(1-R)$, respectively, $R$ being the reflectance of the surface at the wavelength of the scintillation light and $m$ the fraction of specular reflection at the surface. For the simulation, $R$ was regarded as a variable input parameter, while $m$ was set to 0.5 [13]. In the case of photon impact on the PMT window surface, only specular reflection was considered with probability $R_{PM}$, which is the reflectance of the PMT window in contact with the gas. The balance, $(1-R_{PM})$, is assumed to correspond to the transmission of the photon through the window towards the photocathode. The detection of the photon requires its conversion into an electron, process whose average probability is given by the quantum efficiency of the photomultiplier, $Q$, which is taken into account explicitly in Eq. (1), thus outside the simulation. At the liquid-gas interface, it was considered that the photon could undergo either specular reflection or refraction, with probabilities $R_{lg}$ and $(1-R_{lg})$, respectively, $R_{lg}$ being the reflectance of the liquid/gas boundary. Both $R_{PM}$ and $R_{lg}$ were calculated using Fresnel equation [14], which takes into account the dependence of the reflectance on the incidence angle. The refractive index of the xenon vapour in equilibrium with the liquid was approximated to 1 and that of the PMT window, which is made of UV grade fused silica, was taken equal to 1.6 [15, 16]. The refractive index of liquid xenon, $n$, was regarded as a variable input parameter.

When a boundary surface is reached, the competitive processes that can occur at that surface are sampled taking their respective probabilities as listed above. In case of specular reflection and refraction the direction of the photon is changed accordingly. In case of difuse reflection, the direction of the reflected photon was sampled from an angular uniform distribution covering $2\pi$ according to Lambert law [14].

For a given set of values of $n$, $L$, $\eta$, $R$ and $h$, $10^8$ photons were emitted isotropically from random points uniformly distributed over the $\alpha$-source surface and followed until it is lost or is detected by the PMT, as described above. The number of reflections, $r$, and the number of scattering interactions, $s$, undergone by each photon collected at the PMT were counted. The matrix $C$, defined such that the element $C_{rs}$ is equal to the number of photons reaching the PMT photocathode after suffering $r$ reflections and $s$ scattering processes, was computed.

In view of the data analysis, namely the fitting of Eq. (7) to data, the matrix $C$ was computed for the values of $n$ and $L$ forming the rectangular grid in the $(n, L)$-space represented in Fig. 4, and for every value of $h$ at which experimental data were taken. This kind of look-up table of values of the matrix $C$ was computed for the upper limits of the parameters $R$ and $\eta$, which were taken equal to 0.6 and 1, respectively [26]. In so doing, for the values of $h$, $n$ and $L$ mentioned above, the elements of the matrix $C$ allow the analytical calculation of $k$ for any other values of $R$ and $\eta$ in the intervals [0,0.6] and [0,1] respectively, using the relation

$$k(h; n, L, R, \eta) = \left( \sum_{s=0}^{s_{max}} \sum_{r=0}^{r_{max}} \frac{C_{rs}(h; n, L)}{C_{00}(h; n, L)} \left( \frac{R}{R_{max}} \right)^r \eta^s \right) - 1 \qquad (10)$$

During the fitting procedure, the value of $k$ for any set of values of the parameters $(n, L, R, \eta)$ and the variable $h$ was obtained by linear interpolation of the values of $k$ between the closest vertices of the $(n, L)$-grid (see Fig. 4) which, in turn, had been computed for the current values of $R$ and $\eta$ using Eq. (10).

### B. Fitting and Results

As it was referred in Sec. III, two sets of experimental values of $N(h)$ were measured corresponding to the two different positions of the $\alpha$-source, at $H = 187$ mm and $H = 93$ mm. To determine the refractive index and attenuation length of the scintillation photons, Eq. (7) was fitted to the experimental data of both sets by using the method of least squares, i.e., by minimizing the function

$$\chi^2 = \sum_{j=1}^{2} \sum_{i=1}^{M_j} \left( \frac{N_{exp}(h_{ij}) - N(h_{ij})}{\sigma_{ij}} \right)^2 \qquad (11)$$

Here, the index $j$ refers to the dataset ($H_1$=187 mm and $H_2$=93 mm), being $M_j$ the number of points in the $j$ dataset. Hence, $h_{ij}$ is the thickness of the liquid layer above the source (see Fig. 1) at data point $i$ of the dataset $j$, and $N_{exp}(h_{ij})$ is the number of photoelectrons measured with an uncertainty $\sigma_{ij}$. $N(h_{ij})$ was calculated using Eq. (7), with $k(h_{ij})$ computed by Monte-Carlo simulation as described in the previous section. As it was discussed therein, $k$ can depend not only on $n$ and $L$ but also on two additional unknown parameters, $R$ and $\eta$, i.e., the reflectance of the inner surfaces of the chamber and the probability of scattering of photons in the liquid, respectively. Therefore, the fitting was done with six independent adjustable parameters: $n$, $L$, $\eta$, $R$, $A_1$ and $A_2$, being $A_1$ and $A_2$ the multiplicative constants of Eq. (7) for the set of data corresponding to $H_1$ and $H_2$ respectively.

The minimization was carried out using the downhill simplex method [18] as it does not require derivatives. The starting point for defining the initial simplex was chosen randomly in the parameter space. Once the minimization routine has terminated, it was restarted at the point where it claimed to have found a minimum in order to ensure that it was not trapped in a false minimum. The whole fitting procedure was repeated 1000 times with different starting points randomly distributed in the parameter domain (see Table II). Figure 5 and Fig. 6 show the distributions of the parameters and the $\chi^2$ of the fits, respectively, obtained with one and three restarts. From those figures, it is evident the importance of performing more than one restart in each fitting procedure. No significant changes were observed by increasing further the number of restarts. For the best estimate of the parameters one took the set of values of the fit performed with 3 restarts that gave the smallest value of $\chi^2$. That fit is shown in Fig. 3, together with the data points, and the values of the adjusted parameters are listed in Table II.

For the estimation of the confidence limits, the method known as Monte Carlo simulation of synthetic datasets [18] was used. Essentially, the method comprises the following steps. Firstly, from the experimental dataset $(N_{exp}(h_{ij}), \sigma_{ji})$ several new synthetic datasets are generated, such as $N_{syn}(h_{ij}) = N_{exp}(h_{ij}) + \sigma_{ij} G$, where $G$ is a random value, distributed normally with zero mean and dispersion equal to 1. Secondly, for every synthetic dataset the minimization procedure described above was performed giving rise to a set of fitted parameters. The uncertainty in each parameter was calculated as the standard deviation of the distribution of the values of the parameter obtained by performing the fitting over 1000 synthetic datasets. The distributions of the parameter values are shown in Fig. 7 and the uncertainties are listed in Table II. As it should be expected, the mean values of these distributions ($n$=1.69, $L$=362 mm and $R$=0.20) agree, within the uncertainties, with the values of the parameters corresponding to the best fit to the experimental data ($n=1.69$, $L=364$ mm and $R=0.21$).

Figure 8 shows the confidence regions for two parameters of interest jointly, $n$ and $L$, defined by curves of constant $\chi^2$ of the fit to the experimental data. The pairs of values $(n, L)$ obtained by performing the fit to the synthetic data sets are also represented in Fig. 8 by dots. Their fraction inside each confidence region fully agrees with the value predicted on the basis of the $\chi^2$ contours. Moreover, the projections of the curve corresponding to a confidence level equal to 68.3% for one parameter, agree with the standard deviations of the distributions of the parameters $n$ and $L$ obtained by performing the fits to the synthetic data sets (i.e., the distributions plotted in Fig. 7).

The discussion of the absolute values of $n$, $L$ and $\eta$ obtained from the fit will be given in Sec. V. As for $R$, it is consistent with the data on the reflectance of stainless steel in the VUV region that could be found in the literature [17]. The fitted value of the constant $A$ in Eq. (7) was also found to be consistent with the expected one from an estimate assuming $N_0 = E_\alpha/W_{ph} = 3.37 \times 10^5$, $Q = 0.2$ and $T_2 = 0.95$. $E_\alpha$ is the energy of the alpha particles ($E_\alpha = 5.49$ MeV) and $W_{ph}$ the mean energy for producing a photon ($W_{ph} = 16.3$ eV [19, 20]). The value of the quantum efficiency of the photocatode, $Q$, was taken equal to the one quoted by the manufacturer at $\lambda = 178$ nm and $T_2$ was calculated using Fresnel equation with the refraction index of UV grade fused silica and xenon vapour equal to 1.6 and 1, respectively.

It is worth to mention that the condition $k \ll 1$ referred in Sec. II is fulfilled in the present experimental situation. In fact, for the values of the parameters corresponding to the best fit, one has $k \approx 0.03$ for $h > 15$ mm. It increases for smaller values of $h$, attaining its maximum value of 0.1 for $h = 1$ mm.

## V. DISCUSSION

The attenuation length of scintillation light in liquid xenon obtained in this work is very similar to most of the values reported in literature, ranging from 300 mm to 400 mm [9, 21, 22]. An apparent longer attenuation has also been obtained in some measurements where collection of the reflected light was not taken into account [11, 23]. So, in this type of measurements, it is crucial to suppress the collection of reflected light by a proper design of the set-up or to take it into account in the data analysis.

In the present work, the lower limit for the electron lifetime in the liquid measured in a separate chamber was found to be at least $20\,\mu$s, being limited by the sensitivity of the method used for the measurement. This lifetime corresponds to a concentration of oxygen equivalent impurities of about 0.05 ppm, showing that the liquid is very pure at least with respect to electron attaching impurities.

The measurement of a similar value of $L$ by different authors using different purification methods, contrary to the significant improvement achieved in the mean free path with respect to the attachment of free electrons, has led to the suggestion that L was mainly determined by the Rayleigh scattering [9]. Moreover, the observation that the attenuation length of scintillation in liquid argon increases significantly with the addition of small quantities of Kr or Xe, shifting the wavelength of the emitted photons from 128 nm to 147 nm and 174 nm, respectively, has also been cited in support of that hypothesis [9]. Indeed, as the cross section of the Rayleigh scattering process is proportional to $\lambda^{-4}$, the shift to longer wavelength would account for the observed increase of the attenuation length.

The theoretically estimated scattering mean free path of scintillation photons in both liquid xenon and liquid argon doped with Kr and Xe agrees fairly well with the experimental values of the scintillation photon attenuation lengths [10]. However, the hypothesis of the dominant role of Rayleigh scattering was not yet confirmed experimentally.

In the present work, the possibility of occurrence of Rayleigh scattering was introduced in the data analysis and was parameterised in terms of the fitted parameter $\eta$ (see Sec. IV A). According to its definition, $\eta = 0$ means that the only process responsible for the finite value of $L$ is the absorption and, conversely, the upper limit, $\eta = 1$, corresponds to $L$ entirely determined by scattering. From Fig. 7 one can see that the most probable value of $\eta$ is 0. In Fig. 9(a), the $\chi^2$ of the fit to experimental data is plotted as a function of the parameter $\eta$. Figure 9(b) shows similar plot but when the three data points corresponding to the smaller values of $h$ for $H = 193$ mm were not included in the fit. In both figures the horizontal line corresponds to the confidence level of 68.3%. Therefore, we consider that the experimental data do not allow a confident conclusion about the value of $\eta$. It should be stressed that the values of $n$ and $L$ obtained from the fits with and without those three data points are the same.

Regarding the refractive index of liquid xenon for its scintillation, the result obtained in the present work ($n = 1.69 \pm 0.02$) is in good agreement with the value recently calculated ($n = 1.68$) using a free exciton model based on a dispersion relation reported for xenon gas in the UV [24]. Significantly lower value of the refractive index ($n = 1.5655 \pm 0.0024 \pm 0.0078$) has been measured in [5] at xenon triple point. Regretfully, the paper does not provide enough details to enable discussion of possible reasons for the discrepancy. Similar low value, $n = 1.59$ at $\lambda = 175$ nm, was obtained earlier by Braem *et al.* [21] from experimental data [6] at $\lambda > 350$ nm by extrapolation with a non-specified modified form of the Lorentz-Lorenz equation. However, the authors stressed that the extrapolation is very sensitive to the functional form of the extrapolation and to small changes in the values of the refractive index at the longer wavelengths. As an indirect indication of a higher value of $n$, one can refer recent experimental results on the transport of liquid xenon scintillation light [25] that point to a refractive index of liquid xenon higher than that of fused silica, which is 1.6 at 178 nm [15, 16].

## VI. CONCLUSIONS

The refractive index of liquid xenon for its own scintillation at 170 K was measured to be $1.69 \pm 0.02$. This result is in good agreement with $n = 1.68$ obtained using a free exciton model based on a dispersion relation reported for xenon gas in the UV [24].

The attenuation length of the scintillation photons in liquid xenon was also measured. The value obtained, $L = 364 \pm 18$ mm, is comparable with most of the published data. The possibility of occurrence of both absorption and Rayleigh scattering of the photons in the liquid was taken into account in the analysis. However, the data obtained at the present experimental conditions do not allow to conclude about the relative contribution of these processes. A dedicated experiment is needed to clarify this issue.

**Acknowledgments**

This work was financed by the project CERN/P/FIS/1594/1999 from the Fundação para a Ciência e Tecnologia, Portugal. Vladimir Solovov was supported by a fellowship PRAXIS XXI/BD/3892/96 from the same organisation. Akira Hitachi received a fellowship from Fundação Oriente, Portugal.

TABLE I: The boundary processes taken into account in the simulation of light transport in the chamber and their probabilities: $R$ was regarded as unknown quantity, $m$ was set equal to 0.5 (Ref. 13), $R_{PM}$ and $R_{lg}$ were calculated using Fresnel equation.

| Boundary | Processes | Probability |
|---|---|---|
| SS/liquid and SS/gas [a] | Specular reflection | $R * m$ |
| | Diffuse reflection | $R * (1 - m)$ |
| | Absorption | $1 - R$ |
| Gas/PMT window | Specular reflection | $R_{PM}$ |
| | Transmission | $1 - R_{PM}$ |
| Liquid/gas | Specular reflection | $R_{lg}$ |
| | Transmission | $1 - R_{lg}$ |

[a]SS stands for stainless steel

TABLE II: Domain of search, best estimate and uncertainty of the adjusted parameters.

| Parameter | Domain | Best estimate | Uncertainty |
|---|---|---|---|
| n | 1.4...2.0 | 1.69 | 0.02 |
| L (mm) | 150...1000 | 364 | 17.7 |
| R | 0...0.6 | 0.21 | 0.02 |
| $\eta$ | 0...1 | [a] | [a] |

[a]See Sec. V

FIG. 1: Schematic drawing illustrating the principle of the experimental method.

FIG. 2: Schematic drawing of the experimental set-up: 1 - chamber; 2 - PMT; 3,4,5 - floating $\alpha$-source ($^{241}$Am) assembly; 6 - permanent magnet; 7 - diaphragms; 8,9 - level-meter; 10 - alcohol cooled with liquid nitrogen; 11 - thermosensors.

FIG. 3: Number of photoelectrons, as a function of the thickness of the liquid xenon layer, measured at two different positions of the $\alpha$-source (see Fig. 2). The solid line is the best fit of Eq. (7) to data with $n = 1.69$, $L = 364\,\text{mm}$, $R = 0.21$ and $\eta = 0$.

FIG. 4: Grid of points in the $(n, L)$-plane where the matrix $C$ was calculated by Monte-Carlo simulation.

FIG. 5: Distribution of the adjusted parameters obtained by carrying out the fitting with 1000 starting points randomly distributed in the parameter domain. The number of restart points considered in each fitting was equal to one (dashed lines) and three (solid lines).

FIG. 6: Distribution of the $\chi^2$ values of the fits to experimental data obtained with 1000 different starting points randomly distributed in the parameter domain. The number of restarts in each fit was set equal to one (a) and three (b).

FIG. 7: Distributions of the fitted parameters obtained by the method of the synthetic datasets.

FIG. 8: Confidence regions in the $(n, L)$ domain defined by $\chi^2 = const$ contours. Curves $B$ and $C$ delimit confidence regions of 68% and 95%, respectively, for the two parameters jointly. The projections of curve $A$ on the axes define a 68% confidence interval for each parameter separately. The dots correspond to the $(n, L)$-pairs obtained by performing the fit to the synthetic data sets.

FIG. 9: $\chi^2$ of the best fit of Eq. (7) to experimental data as a function of the value parameter $\eta$. (a): with all the data points; (b): without the three data points taken at $H = 187\,\text{mm}$ and with $h \leq 22\,\text{mm}$ (see Fig. 3).

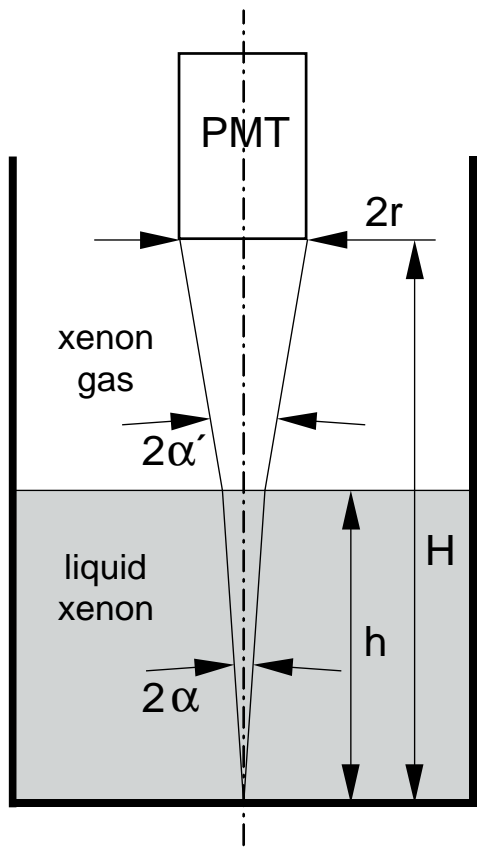

Fig.1

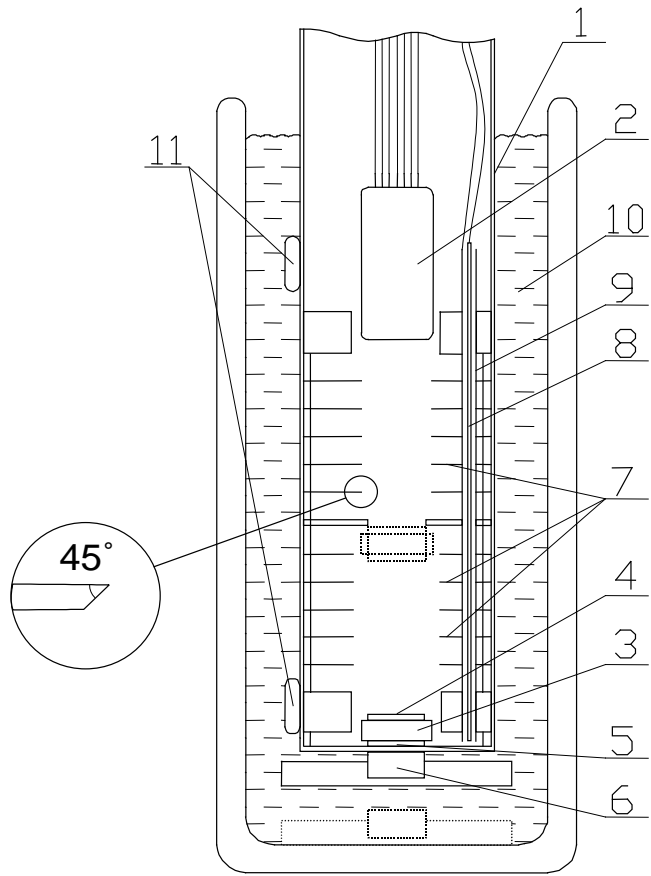

Fig.2

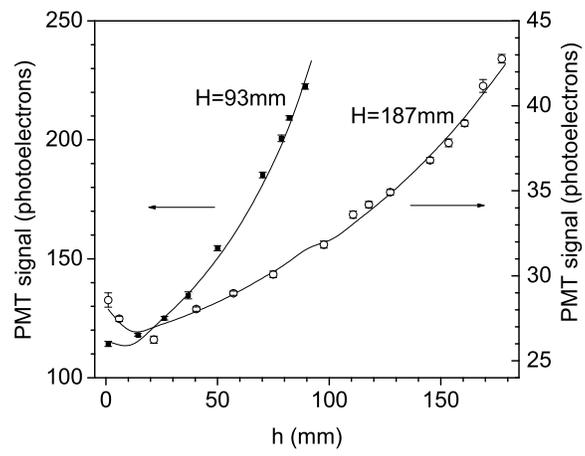

Fig.3

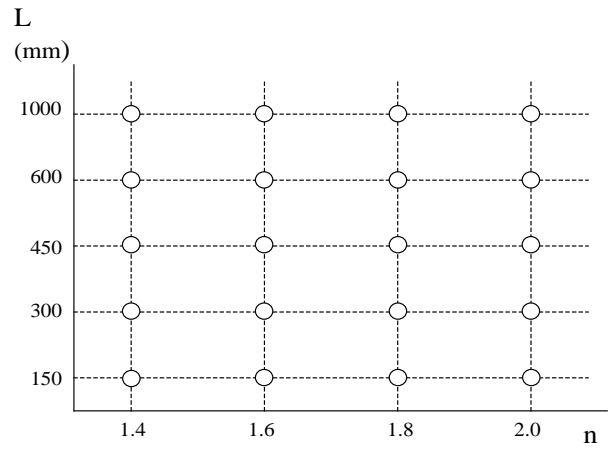

Fig.4

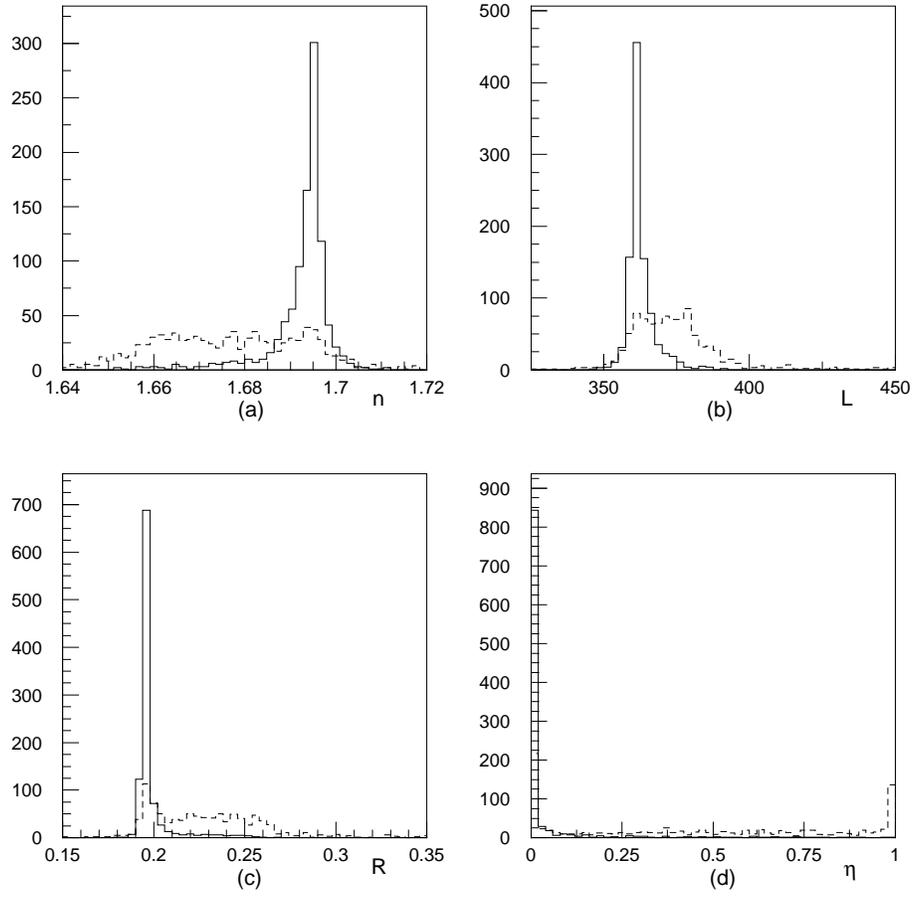

Fig.5

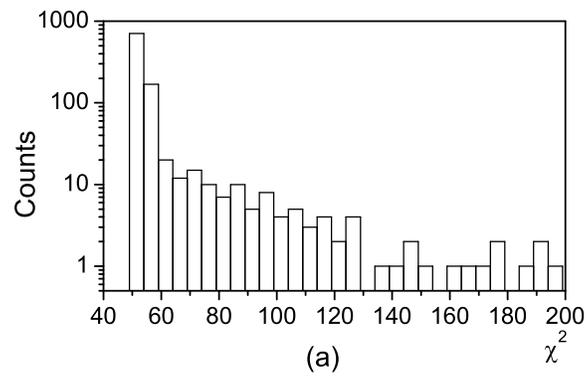

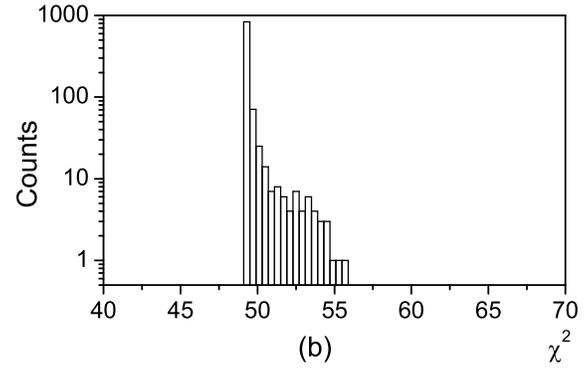

Fig.6

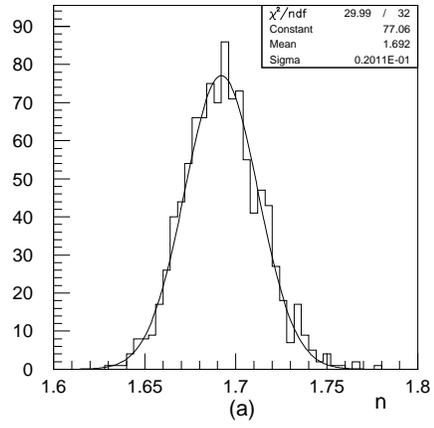
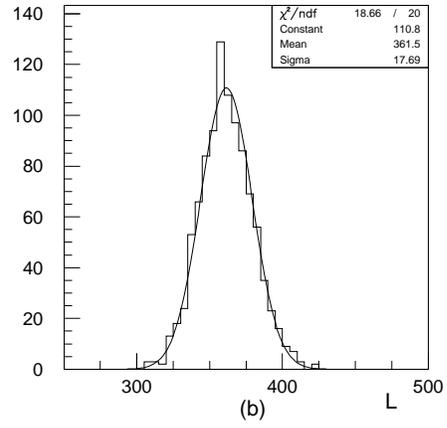
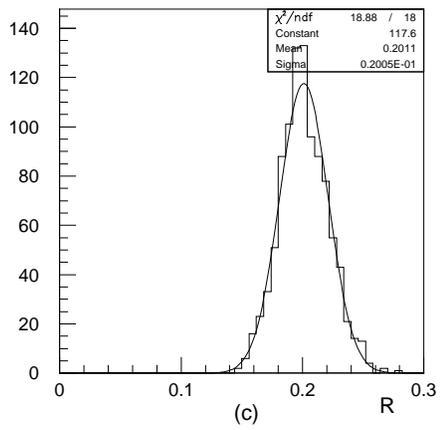
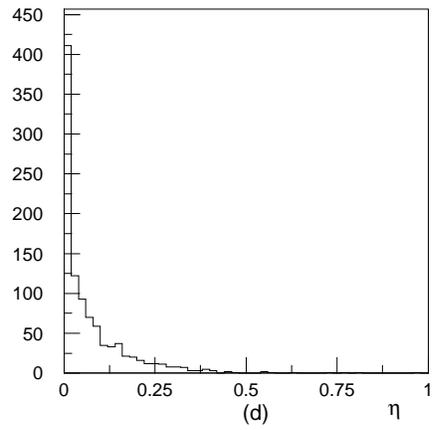

Fig.7

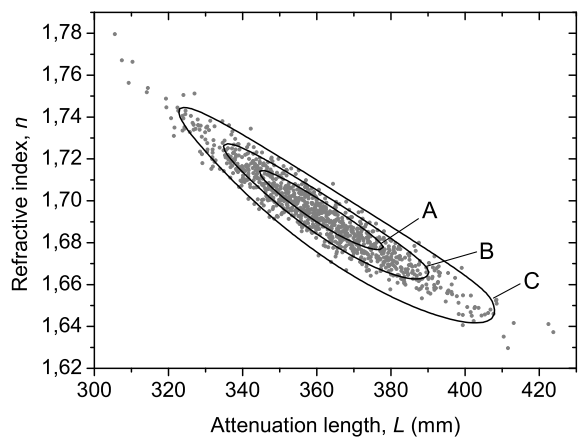

Fig.8

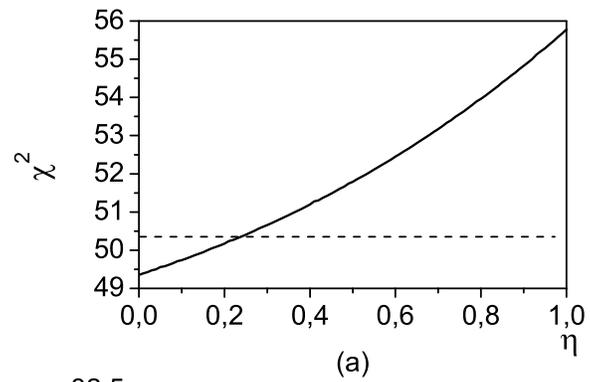

(a)

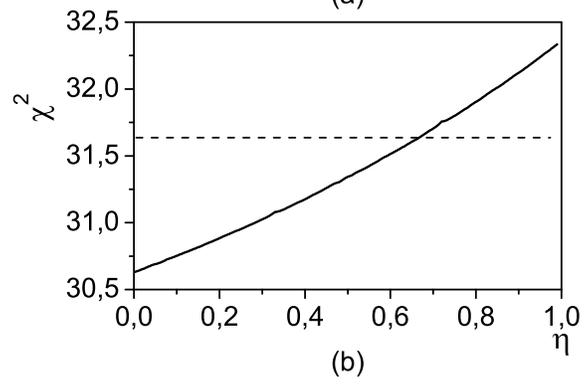

(b)

Fig.9